\documentclass[universe,article,accept,pdftex,moreauthors]{Definitions/mdpi} 
\firstpage{1} 
\pubvolume{9}
\issuenum{1}
\articlenumber{40}
\pubyear{2023}
\copyrightyear{2023}
\externaleditor{Academic Editor: Lorenzo Iorio}
\datereceived{27 November 2022} 
\daterevised{2 January 2023} 
\dateaccepted{4 January 2023} 
\datepublished{8 January 2023} 
\hreflink{https://doi.org/10.3390/ \linebreak universe9010040} % If needed use \linebreak
\Title{Hot Spots in Sgr~A* Accretion Disk: Hydrodynamic Insights}

\TitleCitation{Hot Spots in Sgr~A* Accretion Disk: Hydrodynamic Insights}

\Author{ 
Elizabeth P. Tito $^{1,}$*\orcidB{}, 
Victor P. Goncharov $^{1,2}$\orcidC{} and  
Vadim I. Pavlov $^{1,3}$\orcidA{}
}

\AuthorNames{Elizabeth P. Tito, Victor P. Goncharov, Vadim I. Pavlov}

\AuthorCitation{Tito, E.P.; Goncharov, V.P.; Pavlov, V.I.}
\address{%
$^{1}$ \quad Scientific Advisory Group, Pasadena, CA 91125, USA \\ 
$^{2}$ \quad A. M. Obukhov Institute of Atmospheric Physics RAS, 109017 Moscow, Russia\\
$^{3}$ \quad Facult\'{e} des Sciences et Technologies,  Universit\'{e} de Lille,  
 F-59000 Lille, 
France
}

\corres{Correspondence: eptito@gmail.com} 

\abstract{The recent image of our galaxy's supermassive black hole Sgr~A* derived from the 7 April~2017 data of the Event Horizon Telescope Collaboration shows multiple hot spots in its accretion disk.  Using the analytical framework, we demonstrate that the observed hot spots 
may not be  disjoint elements but causally linked components (``petals'') of one rotating quasi-stationary macro-structure formed in the thermo-vorticial field within the accretion disk. 
}


\begin{document}
%%%%%%%%%%%%%%%%%%%%%%%%%%%%%%%%%%%%%%%%%%
 
%%%%%%%%%%%%%%%%%%%% 
%\part is level -1
%\chapter is level 0
%\section is level 1
%\subsection is level 2
%\subsubsection is level 3
%\paragraph is level 4
%\subparagraph is level 5
%%%%%%%%%%%%%%%%%%%%
% for no numbering whatsoever, use \setcounter{secnumdepth}{-2} or a lower value.
%\renewcommand{\thesubparagraph}{\S\arabic{subparagraph}}

\renewcommand{\thesubparagraph}{(\arabic{subparagraph})}
\setcounter{secnumdepth}{5}
\setcounter{tocdepth}{4}

\section{Introduction}
\label{s:1}

\begin{figure}[H]
\centering
\begin{minipage}{0.34\textwidth}
     \includegraphics[width=0.97\textwidth]{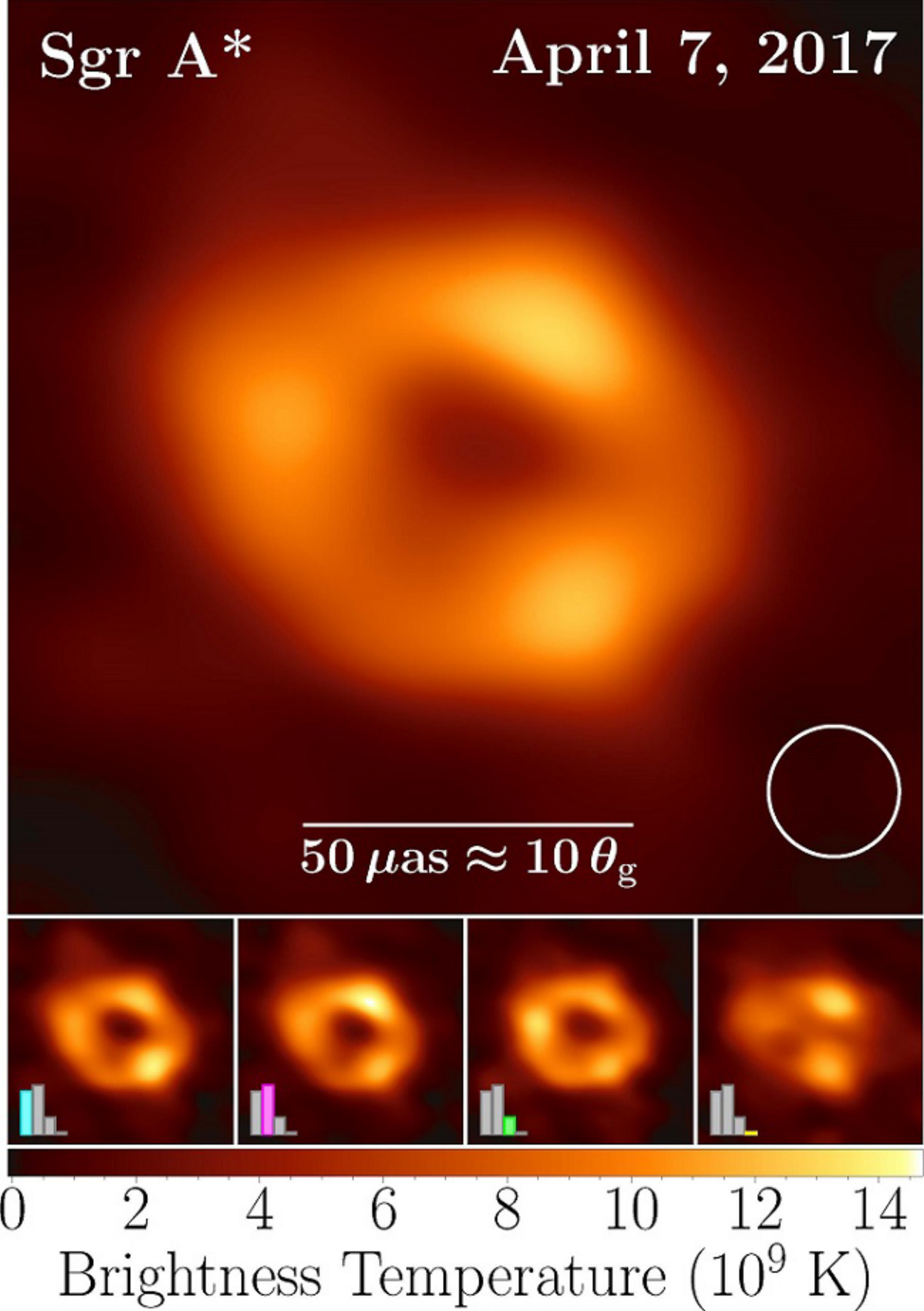}
 \end{minipage}
 \begin{minipage}{0.47\textwidth}
 \centering
        \includegraphics[width=0.97\textwidth]{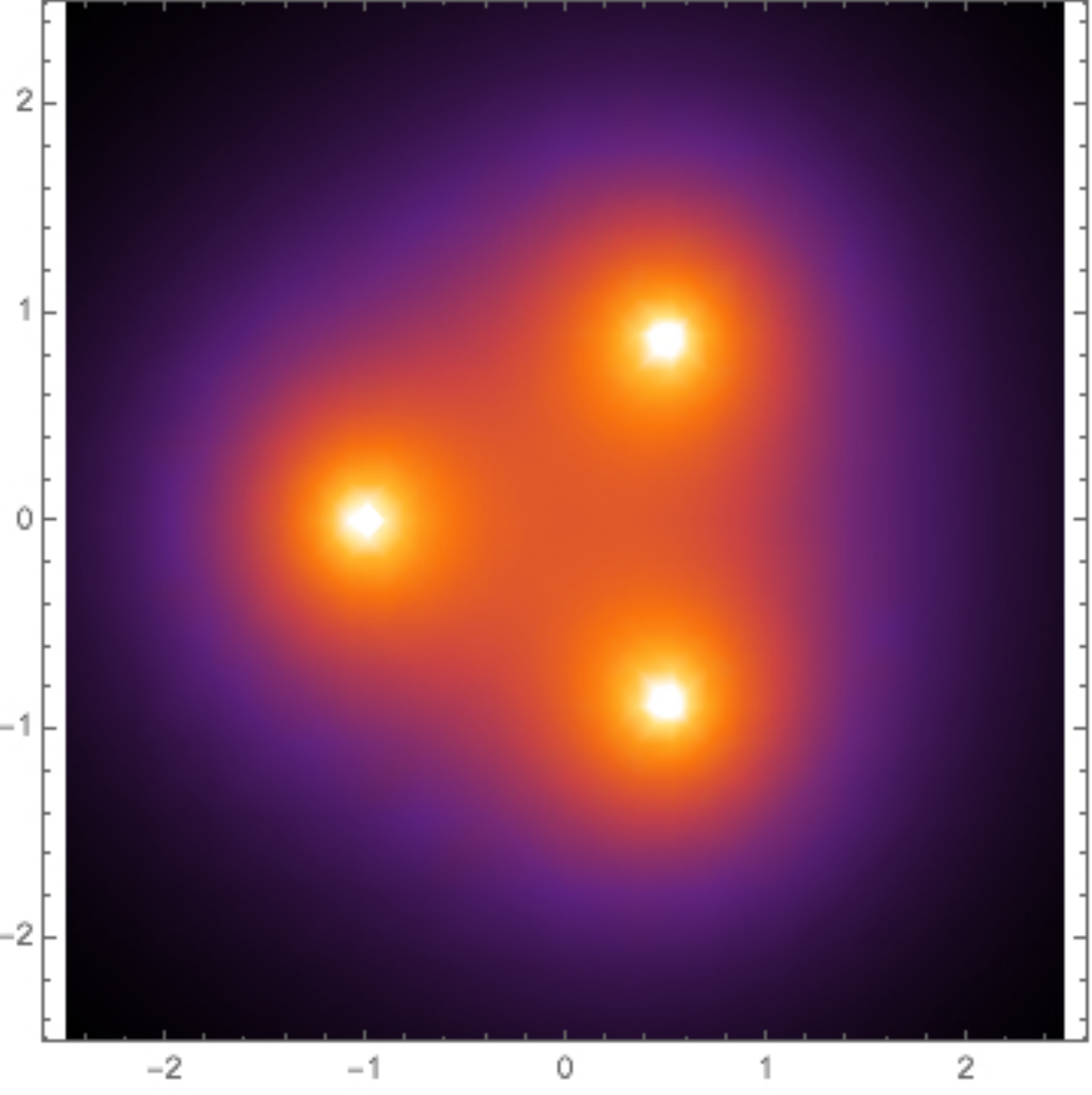}
 \end{minipage}
  \caption{
Left panel~({\bf A}):  
Image of Sgr~A* from Ref.~\cite{coll2022a}. 
Representative EHT image of Sgr~A* from observations on~7 April 2017. 
This image is an average over different reconstruction methodologies
(CLEAN, RML, and Bayesian) and reconstructed morphologies. Color denotes
the specific intensity, shown in units of brightness temperature. The inset circle
shows the restoring beam used for CLEAN image reconstructions ($20~\mu \rm as$ 
FWHM). The bottom panels show average images within subsets with similar
morphologies, with their prevalence indicated by the inset bars. 
 Right panel~({\bf B}): 
  Normalized distribution of temperature-excess 
  in an accretion disk for the model of localized vortices (Section~\ref{s:3}). 
  }      
 \label{Fig1}   
 \end{figure}

The image of our galaxy's central supermassive black hole Sagittarius~A* (Sgr~A*),
derived recently by the Event Horizon Telescope (EHT) Collaboration  
(see Figure~\ref{Fig1}A and Refs.~\cite{coll2022a,coll2022b,coll2022c,coll2022d,coll2022e,coll2022f}) 
shows a multi-spot structure of its accretion disk.  
The disk structure is a product of complex state-of-the-art data analysis rather than a direct observation.  

\begin{figure}[H] 
\includegraphics[width=0.30\textwidth]{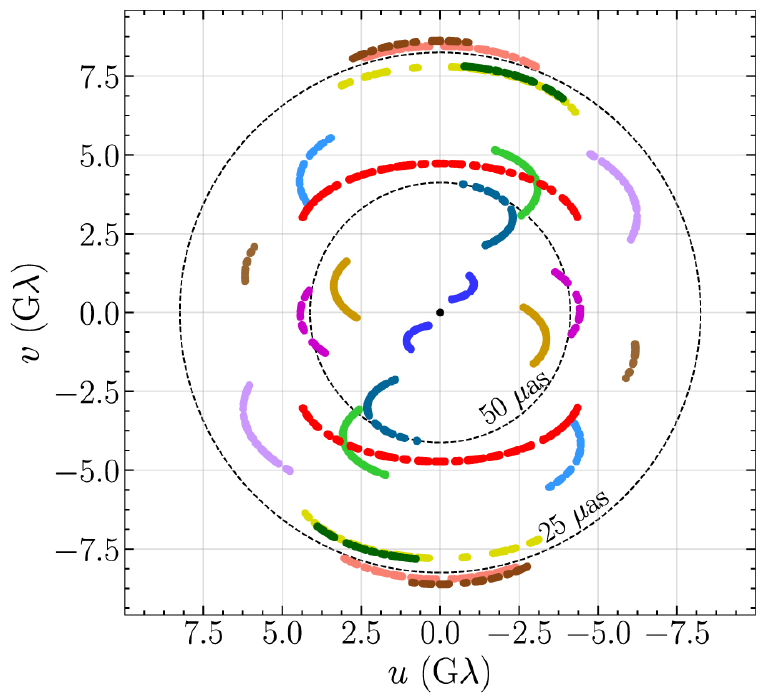}
\caption{
From Ref.~\cite{coll2022a} (one panel from original Figure~2).  
EHT baseline coverage, where dimensionless coordinates $u = (u,v)$ 
give the projected baseline vector for
each antenna pair in units of the observing wavelength. 
}
 \label{Fig2}
\end{figure}

The EHT---a collection of radio-telescopes scattered around the 
Earth---operates in the digital interferometer mode: the signal from each antenna is recorded, and then the image of the object is restored using correlation analysis. 
Sophisticated data-processing algorithms have permitted the EHT to achieve angular resolution on the order of 20~microarcseconds. 
At the level of sensations, this is equivalent to the ability to read newspaper headlines on the Moon. 
However, as Figure~\ref{Fig1}A indicates, this resolution scale is comparable to the size of Sgr~A* itself; the accretion disk is slightly greater ($\sim$$50~\mu \rm as$). 
Furthermore, the EHT telescopes could only record data from a small study area for a short period of time  (see colored zones in Figure~\ref{Fig2}).  Many (white) parts have remained unexplored.  To restore the full mosaic, the algorithms had to fill the gaps. 

The shape of the observed structure (Figure~\ref{Fig1}A)---even 
if the structure is short-lived---appears to indicate that 
it is likely to be not an artifact of image-reconstruction algorithms, 
but a real phenomenon. 
Using the {\em analytical} framework, we demonstrate that the observed hot spots 
may be not disjoint but 
 causally linked components (``petals'') of 
one rotating quasi-stationary macro-structure 
formed in the thermo-vorticial field within the accretion disk. 

Indeed, 
when a black hole's accretion disk---whose rotation axis is perpendicular to the disk plane---is 
heated non-homogeneously (so temperatures are higher near the outer edge of the disk), 
then, 
in the field of the centrifugal force,   
 spontaneously self-formed hot ``bubbles''
(composed of locally clustered plasma with temperatures in excess of the ``average'', hence with lower densities) 
should move towards the axis of the disk rotation. 
However, when the hot ``bubbles'' are also vortices, 
then each such vortex 
(via the induced velocity field) 
``forces'' 
other vortices to rotate around itself, 
hence diverting their motion ``sideways'', curtailing the movement towards the central axis of accretion-disk rotation.
All these vortices
are subject to the influence of the cumulative velocity field 
induced by all other vortices. 
Thus, 
the radial motion of the vortices towards the axis becomes suppressed. 
As a result, 
if stabilized, 
the vortices take positions equidistantly from the axis and 
self-organize into a symmetric 
thermo-vorticial 
macro-structure 
that rotates as a whole around the mutual center 
 (like in Figure~\ref{Fig1}B).  
The dynamics and longevity of this structure are linked to the thermal and vortical properties of the system and its elements. 
Visually, if observed, the petals of this structure look like bright ``hot spots''.

In this paper, we  
consider the EHT image 
from the perspective of theoretical hydrodynamics. 
In particular, 
we describe a model of  large-scale stationary rotating 
heated vortices. 
However, 
we consider not the usual hydrodynamic field of vorticity but a complex thermo-hydrodynamic field system that under certain circumstances may self-organize  into regular structures.  
The physical and mathematical underpinnings of this analytical approach are elaborated in the references provided in the relevant places. The explanation of their details is beyond the scope of this paper.
To avoid any confusion, let us also emphasize upfront that 
we work with the {\em field}, not with individual particles (their trajectories or orbits). Perhaps what may help the reader grasp this nuance better is the reminder that the {\em velocity} of displacement of electrons in a usual house wire is not the same thing as the {\em speed} of propagation of the electro-magnetic field perturbation along that same wire. 
As the result of our analysis, we 
show that multi-hot-spot thermo-vorticial macro-structures 
may 
indeed 
self-organize 
in the accretion disk. 
The model makes it possible to determine 
basic characteristics of 
such structures, 
for example, 
the horizontal space-scale, 
the period of proper rotation, 
and the peak temperature magnitude in the vortex. 

The paper is organized as follows: 
Section~\ref{s:2} presents the model,   
Section~\ref{s:3} presents the results, 
and 
Section~\ref{s:4} summarizes the conclusions.

\section{Model} 
\label{s:2}

The model setup is straightforward: 
a black hole pulls in and crushes the matter from the surrounding space;  
the particles are then accelerated to 
near-light velocities and twisted around the black hole, forming a flattened accretion plasma 
disk in the equatorial plane. 

We will use the spacetime metric entirely characterized by the black hole mass parameter 
and its “spin” (described in our Refs.~\cite{Tito2018,Tito2021}; 
for more details, 
see also 
\cite{Landau_Fields,Misner1973,Shapiro1983,Visser2007,Frolov2011,Weinberg1972}, and bibliographies therein). 
We will assume that
the mass of the accretion disk is negligible compared to the black hole ``mass''~$M$, probably $(10^{-5} \div 10^{-4}) M_{\odot}$;  
the radiative cooling does not strongly affect the dynamics of fluid motion;  
and 
the electrons and ions are very weakly coupled by Coulomb interaction and  
hence ions and electrons plasmas components have different temperatures, 
$T_e \gg T_i$ 
(see Ref.~\cite{Kadomtsev1988}), and thus 
it is the 
electron component that contributes the most to 
the equation of state of the accretion disk matter. 
Due to the large difference in the masses of electrons and protons, electrons are highly mobile and provide quasi-neutrality of the plasma. Due to the high conductivity of the plasma, its own magnetic field can be considered as a field ``frozen'' into the plasma. 

Generally speaking, equations of fluid motion in the vicinity of a black hole must be written using the concept of relativistic dynamics. 
The key points are as follows: 
We suppose that the space-time near the (non-charged) black hole Sgr~A* is described by the Kerr metric---an exact, singular, stationary, and axially symmetric solution of the Einstein–Hilbert equations of the gravitational “field” in vacuum. Next, we introduce the Boyer–Lindquist 4-coordinates, $q^{\alpha} = (t, r, \theta, \phi)$ (it is well known that besides the Boyer–Lindquist coordinate representation, other representations of space-time locations exist). In terms of the Boyer–Lindquist coordinates, the square of interval is written as 
$d s^2 = g_{\alpha \beta} (r / r_g , \theta) dq^{\alpha} dq^{\beta}$ 
with $\alpha, \beta = 0, 1, 2, 3$, i.e., the components of $g_{\alpha \beta}$ depend only on the dimensionless combination $r_g / r$ and $\theta$. 
Here, $r_g = 2 G M / c^2$ is the Schwarzschild radius, $c$ is the speed of light, $G$ is the gravitational constant, and $M$ is the “mass” of the black hole. The off-diagonal term $g^{03}$ in the metric tensor is proportional to the rate of the black hole's own rotation and to $1/r$. 
For the Minkowski tensor,
we use the metric signature $diag(+ - - -)$ (see, for example, Ref.~\cite{Tito2021},  
and Refs. therein). 
To satisfy the principle of causality for moving material objects,  
obviously, $d s^2 > 0$. 
The four time-space coordinates $q^{\alpha} = (t, r, \theta, \phi) $ give the location of a world-event from the viewpoint of a remote observer. The meaning of space coordinates $r, \theta, \phi $ is clear once transitioned to the limit $r \gg r_g , \, r \gg \omega r_g^2 /c $. When the square of the interval becomes $ds^2 \rightarrow c^2 d t^2 - d r^2 - r^2 (d \theta^2 + sin^2 \theta ~ d \phi^2)$, i.e., at infinity, parameters $r, \theta , \phi$ may be interpreted as the standard spherical coordinates in flat space-time. As for the parameter $r$, strictly speaking, note that it is not the “distance” in the usual meaning from the center of black hole. This is because, for any material object, 
in the space-time defined by equation  $d s^2 = g_{\alpha \beta} dq^{\alpha} dq^{\beta}$,   
no central point $r = 0$ exists in the sense of a world-event on a valid world-line.

Next, we consider the motion of the medium far away from the event horizon, 
i.e., when 
parameter $r$ is meaningfully greater than $r_g$.
For the flow at $r  > 3 r_g$, 
in the expansion of the metric tensor, 
we may neglect the terms 
of order $(r_g/r)^2$ and greater. 
They contribute less than $(1/3)^2 \sim 10\%$ to the components of the metric tensor. 
Such omission of smaller terms makes our approximation 
Newtonian or 
post-Newtonian.  
The non-diagonal metric-tensor term 
(describing 
involvement of the medium in the rotation of space-time in the vicinity of the black hole)  
gives rise to 
the ``force'' analogous to the traditional Coriolis force 
in the equations for medium flow.  

Hence, 
we write the system of equations of relativistic fluid dynamics 
in the curved space-time  and expand the
metric tensor and the fluid energy--momentum tensor 
into a series with respect to small parameters $ r_g / r  <  1$ and $\sim c^{-1}$. 
We keep only the leading terms in the equations of fluid motion. 

Next, the fluid is presumed to be localized near surface $\theta = \pi/2$ (in a pancake-like accretion disk)---the flows of the disk medium are considered only near this surface. 
Then, we can transition to cylindrical coordinates $q^i = (r, \phi ,z)$ and 
presume that the gas particles orbit near the  $z = 0$ plane 
and the vertical component of their velocity $v_z \ll max ( v_r, v_\phi )$. 

In this model, 
when $r$ exceeds the Schwarzschild radius $r_g$ (specifically,  $r > 3r_g$), then 
the vertical component of gravity acceleration 
$g_z =$  $-~G~M~z/(r^2~+~z^2)^{3/2}$ $\equiv - K^2 z (1 + z^2/r^2)^{-3/2}$ 
is balanced by the pressure gradient. 
Here, 
$G$ is the gravitational constant, 
$c$ is the speed of light, 
$K^2 = G M / r^3 = c^2 r_g / (2 r^3)$, 
and $K \sim r^{-3/2}$ is the radius depending so-called Kepler parameter: 
 the angular velocity of a test particle in a circular orbit at distance $r$ from a point mass $M$ in Newtonian approximation of gravity. 

Gas pressure along the direction 
perpendicular to the disk plane $(x,y)$  
is determined by the  hydrostatic equilibrium,
$ d P = \rho g_z d z $. 
Generally speaking, pressure may  
have important contributions from  electromagnetic radiation and induced magnetic field. 
The simplest case, however, is when the pressure is dominated by gas pressure, and the vertical temperature distribution is isothermal, 
which is roughly appropriate when the disk is optically thick and 
externally heated. 
The equation of state of gas/plasma is then $P = s^2 \rho$, where  
$s$ is the ``isothermal sound speed'' (which is 
not a function of the transversal coordinate $z$). 

If we further assume that $z \ll r$, the equation of hydrostatic equilibrium becomes  
$s^2 d \rho = - s^2 h^{-2} \rho z d z $,   
the solution of which is $\rho (z) = \rho_0 \exp ( - z^2 / 2 h^2 ) $. 
Here, the transversal space scale $h$ is introduced, $h = s \, r^{3/2} / (G M)^{1/2}$. 
Parameter  $\rho_0$ has the meaning of density at the disk mid-plane (at $z = 0$), and 
parameter $h$ is the characteristic local thickness of the accretion disk. 

In order for the accretion disk to be considered thin, it is necessary that 
$h / r \simeq s r^{1/2}/ (G M)^{1/2}  \sim (s / c) (r / r_g)^{1/2} \ll 1$. 
On the other hand, for great distances away from the black hole ($r > r_g$),  specific relativistic effects may be neglected or parametrized. 
Thus, we obtain 
natural bounds: 
$r_g < r < r_g (c / s)^2$. 

{\bf Equations of Motion:} 
The equations of motion for an inviscid incompressible fluid express the laws of conservation for mass, momentum, and energy.  
The flow of gas may be considered   incompressible  
(see, for example, \cite{LandauLifshitz1987})  
when its 
velocity $v$ satisfies condition $v^2 \ll s^2$ and 
the characteristic time scale of the flow change 
$T \gg L / s$ 
where $L$ is the characteristic spatial scale where 
the flow characteristics change substantially. 
In this case, the mass conservation law is expressed as 
$div \, {\mathbf v} = 0$, i.e., the law is not $\rho = Const$, but $\partial_t \rho + v_j \partial_j \rho = 0$. 

We assume that the domain of the disk where the vortex structure of interest is formed, is characterized by an approximately constant angular velocity  $\mathbf \Omega$. In a 
coordinate system rotating with the angular velocity 
$\mathbf \Omega = (0, 0, \Omega)$,  where  
$x \equiv x_1$ and $y \equiv x_2$ are the axes in the horizontal plane and $z$ is vertically upwards, the basic  
equations of motion become
\begin{eqnarray}
\partial_j v_j = 0 ,  \label{eq:1}\\
\rho (\partial_t v_i + v_j \partial_j v_i + 2 \epsilon_{ijk} \Omega_j v_k ) = - \partial_i P - \rho \partial_i \Phi , \label{eq:2}\\ 
\partial_t \theta + v_j \partial_j \theta = 0.\label{eq:3}
\end{eqnarray}
Here,  
$v_i$ are the components of the velocity field, 
$\rho$ is density,
 $P$ is pressure, and
$\theta$ is temperature. 
The potential $\Phi$  of the force field is  
$\Phi \simeq - K^2 r^2 + (1/2) K^2 z^2 - (1/2) | [{\mathbf \Omega}, {\mathbf r}] |^2$. 
Furthermore, here, 
$[ {\mathbf a}, {\mathbf b} ]$ is the cross product of  vectors $\mathbf a$ and $\mathbf b$;  
and 
$\epsilon_{ijk}$ is the alternating tensor ($\epsilon_{123}=1$, zero for any two indices being equal, $+1$ for any even number of permutations from $\epsilon_{123}$, and $-1$ for any odd number of permutations). 

Due to the assumption of  an incompressible medium (implying $s^{-2} \rightarrow 0$), 
 the equation of state $\rho = \rho (\theta , P)$  turns into the density expression, which only depends on the temperature (and not on the pressure).
In fact, $\nabla \rho = (\partial \rho / \partial \theta) \nabla \theta + (\partial \rho / \partial P) \nabla P \simeq - \rho \beta \nabla \theta$.  
The coefficient of thermal expansion, $\beta = - \rho^{-1} (\partial \rho / \partial \theta)$, may be assumed constant and positive.  (For gases, $\beta=1/\theta_0$.) Thus, we can set $\rho = \rho_0( 1- \beta (\theta - \theta_0)  )$, where subscript zero denotes the reference values. 
Due to the fact that $\beta (\theta - \theta_0)$  is generally significantly less then one, 
one may neglect the density variations in all principal terms and hence replace $\rho$ with the constant value $\rho_0$, except in the ``buoyancy'' term, which is proportional to $\partial_j \Phi$ (see,  for example, \cite{LandauLifshitz1987,Turner1979}).

Next, we apply the curl operator ${\mathbf \nabla} \times$ 
to the linear momentum conservation \mbox{Equation~(\ref{eq:2}).} This gives 
\begin{eqnarray}
\bigg[ \nabla - \beta \nabla \theta , \frac{d {\mathbf v}}{d t} + 2 \bigg[ {\mathbf \Omega}, {\mathbf v}  \bigg]  \bigg] = \beta \bigg[ \nabla ( \tau_s + \tau ), \nabla \Phi \bigg] \, , 
\label{eq:4}
\end{eqnarray}
where brackets symbolize cross-product. 
Since 
$\beta | \nabla \theta | \ll 1$,  term 
 $\beta [ \nabla \theta, {\mathbf A}]$ is small with respect to $[ \nabla , {\mathbf A}] $ at the horizontal scales 
typical
 for any hydrodynamical vector $\mathbf A$ and may be not taken into account. 
 
Consider now temperature $\theta$ as $\theta = \theta_0 + \tau_s + \tau$, where 
$\theta_0$ is the (constant) baseline temperature of the accretion disk, 
quantity $\tau_s$ is an axially symmetrical part of the temperature distribution that is not time-dependent, 
and  $\tau = \tau (t, x_j )$ 
is the dynamical quantity related to the vortex structures in the fluid. 
When 
${\mathbf r} = ({\mathbf x}, z), v_j = ({\mathbf v}, w)$, 
then $\omega_i = \epsilon_{ijk} \partial_j v_k + 2 \Omega_i$, 
and Equation~(\ref{eq:4}) can be rewritten in the tensorial form as 
\begin{eqnarray}
{( D_t + w \partial_3) \omega_i} = \omega_j \partial_j v_i + \beta \epsilon_{ijk} ( \partial_j ( \tau_s + \tau ) ) \partial_k  \Phi , 
\label{eq:5}
\end{eqnarray}
where $D_t = \partial_t + ({\mathbf v} \cdot \nabla)$ is the substantial derivative,  
${\mathbf v} = (v_1, v_2)$ is the flow velocity, 
and $\nabla = (\partial_1 , \partial_2)$ is the gradient operator with components  $\partial_i$ in the ($x,y$) plane. 

In the simplest model---in which the vortex structures are realized in a ”thin” flat sheet of an incompressible inviscid fluid (i.e., when $h \ll L$, $V^2 \ll  s^2 \ll c^2$, $Re \gg 1$)---the $z$-component of velocity, $w$, 
vanishes and may be dropped 
in all formulas  (see, \mbox{for example, \cite{GonPav1998b}}).  
Note also that $\epsilon_{3jk} \partial_j \tau_s \partial_k \Phi \equiv 0$ because of an axial symmetry of both $\tau_s$ and $\Phi$.  
Thus, 
the set of equations for the $z$  component of vorticity $\omega_3$ and for variation of temperature $\tau$ becomes 
\begin{eqnarray}
\partial_j v_j = 0 ,  \label{eq:6}\\
D_t \omega_3 = \beta \epsilon_{3jk} \partial_j \tau \partial_k \Phi , \label{eq:7}\\ 
D_t \tau = - v_i \partial_i  \tau_s,  \label{eq:8}
\end{eqnarray}
where indices $j, k = 1,2$. 
Equation~(\ref{eq:6}) permits the introduction of the stream function  $\psi$:  
\begin{eqnarray}
v_i = \epsilon_{3ij} \partial_j \psi \equiv   \epsilon_{ij} \partial_j \psi  ,     \label{eq:9}
\end{eqnarray}
where tensor $ \epsilon_{ij} $ is the antisymmetric unit tensor of the second order, $\epsilon_{12} = - \epsilon_{21}$, and diagonal components are zero. 
In this case,  vorticity $\omega_3 = - \Delta \psi + 2 \Omega_3$ with $\Delta = \partial_1^2 + \partial_2^2$. 

{\bf Temperature Stratification:} 
To describe the effect of  temperature stratification---i.e., 
to show how stationary temperature increases 
as the distance from the axis of rotation increases---we express 
(using $r = | {\mathbf x} |, {\mathbf x} = (x_1, x_2)$ and Taylor series expansion)  
the background distribution of the temperature $\tau_s$ in the form 
\begin{eqnarray}
\tau_s =  \frac{\alpha}{2} r^2 \; , \;  \alpha > 0 \, ,
  \label{eq:10}
\end{eqnarray}
where  $\alpha$ is the parameter characterizing the ``rapidity'' of increase in temperature with distance from the disk rotation axis. 
(Generally speaking, the question of what shape the temperature profile takes within a black hole's accretion disk is an open one.  Experimental measurements remain challenging, despite significant progress.  See, for example, \cite{Runge2021,Wong2011}.) 
When the leading contribution to potential $\Phi$ comes from the centrifugal effects, we can 
express 
$\Phi = - K^2 |{\mathbf x}|^2 - (1/2) [{\mathbf \Omega}, {\mathbf x}]^2 \simeq - (1/2) [{\mathbf \Omega}, {\mathbf x}]^2$. 
Thus, 
when $\Omega$ is presumed constant in a band of $r$ where the vortex hotspots are forming, 
the set of coupled nonlinear evolution equations becomes
\begin{eqnarray}
\partial_t  \Delta \psi+ \epsilon_{i k} \partial_i \psi \partial_k   \Delta \psi   = \beta \Omega^2 \epsilon_{ik} x_i \partial_k \tau , \label{eq:11}\\ 
\partial_t \tau + \epsilon_{i k} \partial_i \psi \partial_k \tau = \alpha \epsilon_{i k}  x_i \partial_k  \psi .  
\label{eq:12}
\end{eqnarray}
For a general case, 
$\Delta \psi$ should be replaced:  
$\Delta \psi  \rightarrow \Delta \psi - 2 \Omega $. 

Equations~(\ref{eq:11}) and  (\ref{eq:12}) have a transparent physical meaning. 
Their left sides describe transport of dynamic quantities: vortex $\Delta \psi$ and temperature perturbation $\tau$. 
Their right sides describe ``sources'' that generate the vortices and temperature perturbations. 

In other words, the 
vortices are generated by the source  (the right part of  Equation~(\ref{eq:11}) 
which is effective 
(non-zero) only when there exists an inhomogeneous gravity-like force field $\Phi \sim \Omega^2$ with which  temperature perturbation $\tau$ interacts via the equation of state (when $\beta \neq 0$). 
The quantity $\psi$---which characterizes the vortex field in the fluid---is transported by the self-induced flow (the left part of  Equation~(\ref{eq:11})).  
On the other hand, in Equation~(\ref{eq:12}), 
this temperature perturbation $\tau$ is transported by the self-generated flow 
($\psi \neq 0$); 
the temperature perturbation $\tau (t, {\mathbf x})$ is generated by the source  (the right part of  Equation~(\ref{eq:12})), which is non-zero only when  
there exists a spatial and time-independent temperature gradient 
(i.e., when $\tau_s \neq 0$). 
The processes are interlinked because the ``source'' of one dynamic quantity depends on the complex combination involving another dynamic quantity. 
The set of Equations~(\ref{eq:11}) and  (\ref{eq:12}) shows that when 
temperature stratification is absent ($\alpha = 0$) and there is no disk rotation (disk $\Omega = 0$, i.e., centrifugal force is zero), 
then Equations~(\ref{eq:11}) and  (\ref{eq:12}) degenerate into the traditional equations for vortex evolution and transport in a two-dimensional ideal fluid.

{\bf Linear Approximation:} 
Assuming that an excess of temperature $\tau$ above the basic level of temperature is not too large, 
in view of the link  between fields $\tau$ and $\psi$ via \mbox{Equations~(\ref{eq:12}), }
we consider a simple linear dependence between the excess of temperature $\tau$ 
and the vorticity $\psi$ that generates this excess: 
\begin{eqnarray}
 \tau = - C \psi      \, .
 \label{eq:13}
\end{eqnarray} 
Obviously, it follows from  Equation (\ref{eq:13}) and 
from the meaning of quantities $\tau$ and $\psi$ that the dimension of the coefficient of proportionality is 
$[C] = \theta L^{-2} T \equiv [temperature] \times [lengh]^{-2} \times [time]$.  
By substituting Equation~(\ref{eq:13}) into Equation~(\ref{eq:12}), we obtain 
\begin{eqnarray}
\partial_t  \Delta \psi+ \epsilon_{i k} \partial_i \psi \partial_k   \Delta \psi   = - C \beta \Omega^2 \epsilon_{ik} x_i \partial_k \psi , \label{eq:14}\\ 
- C \frac{\beta}{\alpha} \Omega^2 \times C ( \partial_t \psi + \epsilon_{i k} \partial_i \psi \partial_k \psi ) = C \frac{\beta}{\alpha} \Omega^2 \times \alpha \epsilon_{i k}  x_i \partial_k  \psi .  
\label{eq:15}
\end{eqnarray}
Hence, 
we conclude  that both the first and second equations in Equations~(\ref{eq:14}) and (\ref{eq:15}) describe the evolution of the same physical quantity. The equations will be consistent when the following condition is imposed on the current function 
$\psi$: 
\begin{eqnarray} 
( \partial_t +  \epsilon_{i k}  \partial_i \psi  \partial_k  ) ( \Delta \psi -R^{-2} \psi )  = 0 \, .  
 \label{eq:16}
\end{eqnarray}
Here,  parameter $R^{-2} = C^2 ({\beta}/{\alpha}) \Omega^2  $. The dimension of this quantity is $ ( \theta^1 L^{-2} T^1 )^2 \times \theta^{-1} L^2 \times \theta^{-1} \times T^2 = L^{-2}$;
i.e., $R$ is a space scale factor.  
Quantity  $q = \Delta \psi - R^{-2}  \psi$ describes the distribution of generalized vorticity. Its change in time and in space has to satisfy the evolution equation Equation~(\ref{eq:16}).  

The stream function $\psi$ is found via the Green function approach. 
In the symbolic integral form, in the boundless space, it is
\begin{eqnarray}
\psi  ({\mathbf x}, t) = \bigg( ( \Delta - R^{-2} )^{-1} ({\mathbf x}, {\mathbf x}') \bigg)   q ({\mathbf x}', t)  \, .
\label{eq:17} 
\end{eqnarray} 
The subsequent calculation procedure is as follows: 
(i) The initial vorticity distribution is set; the stream function $\psi$ is found from Equation~(\ref{eq:17}); 
together with it, the non-linear evolution Equation~(\ref{eq:16}) is numerically solved. 
(ii) The distribution of vorticity is set in the form of a macrostructure with petals (with constant vorticity inside) whose moving boundaries evolve according to Equation~(\ref{eq:16}); i.e., we are considering 
a region bounded by some closed contour (with possibly a rather complex shape) such that quantity $q ({\mathbf x}, t)$ takes a constant value inside and zero outside. 
For stationary dynamical regimes, this can be accomplished analytically 
(details of the contour dynamics method and of the operator techniques can be found, for example, in 
Refs.~\cite{GonPav1993,GonPav2008}), as well as Refs.~\cite{GonPav1998b,GonPav2000}); 
(iii) For a strongly localized vortex, 
quantity 
$q ({\mathbf x}, t) $ can be parameterized by the function in form $F ({\mathbf x} - {\mathbf x}_0 (t))$, i.e., 
the one with the center at 
coordinate ${\mathbf x} =  {\mathbf x}_0 (t)$, which satisfies the equation of transport $\dot  x_{0 i} (t) + \epsilon_{i k}  \partial_i \psi  = 0$, i.e., when the center of vortex moves according 
to $\dot  {\mathbf x}_{0} (t) - {\mathbf vs. }  [{\mathbf x}_{0} (t)] = 0$. 
(Here, the symbol ``dot'' signifies the derivative with respect \mbox{to time}.)

{\bf Stationary Vortex Structures:} 
Stationary vortex structures---the ones rotating with constant angular velocity 
$\omega$---are simpler to consider 
in a  rotating coordinate system where the structures appear immovable; i.e., 
when rotation direction is co-aligned with $z$ axis,  
the derivative with respect to time becomes 
$\partial_t = - \omega \epsilon_{ik} x_i \partial_k$.  
(The procedure is laid out, for example, in Ref.~\cite{GonGrPav2002}.) 
Indeed, when $f = f(\rho, \phi - \omega t$), 
then the calculation relying on the properties of Jacobeans produces 
$\partial_t f = - \omega \partial_{\phi} f = - \omega \partial (f, \rho) / \partial (\phi, \rho) =  - \omega (\partial (f, \rho) / \partial (x_1, x_2)  )( \partial (x_1, x_2) / \partial (\phi, \rho)) = - \omega (\epsilon_{i k} \rho^{-1} x_i \partial_k f ) (- \rho) $. 
Then, Equation~(\ref{eq:12}) may be rewritten as:  
\begin{eqnarray}
 \epsilon_{i k} \partial_i \bigg( - \frac{\omega}{2} |{\mathbf x}|^2 + \psi \bigg) \partial_k  \bigg( \tau + \frac{\alpha}{2} |{\mathbf x}|^2 \bigg)  = 0 \, ,
 \label{eq:18}
\end{eqnarray}
which is satisfied by the ansatz   
\begin{eqnarray}
 \tau + \frac{\alpha}{2} |{\mathbf x}|^2  = F (- \frac{\omega}{2} |{\mathbf x}|^2  + \psi).    
 \label{eq:19}
\end{eqnarray}
The explicit expression of function~$F$ 
can be found from the obvious fact that temperature-driven flow perturbations must vanish when temperature perturbations vanish themselves. 
The suitable expressions is
function $F (u) = - \alpha u / \omega $. 
Then, temperature fluctuations are expressed via the stream function 
\begin{eqnarray}
 \tau = - \frac{\alpha}{\omega}  \psi      \, .
 \label{eq:20}
\end{eqnarray} 
Combining this with Equation~(\ref{eq:11}) where 
$\partial_t = - \omega \epsilon_{ik} x_i \partial_k$ is taken into account, we obtain the second evolution equation: the one that describes rotation of a stationary vortex structure with angular velocity 
$\omega$ caused by the self-induced field of hydrodynamical velocity:   
\begin{eqnarray}  
  \epsilon_{i k} ( -\omega x_i  + \partial_i \psi ) \partial_k  \bigg( \Delta \psi - R^{-2} \psi\bigg)  = 0 \, .   
 \label{eq:21}
\end{eqnarray}

Once Equation~(\ref{eq:20}) is written out, parameter $R^2$ becomes 
$R^2 = (\alpha \beta)^{-1} (\omega / \Omega)^2 $. 
Here, $\beta$ is the coefficient of thermal expansion. 
For almost all physically realizable situations, 
$\beta > 0$ (the well-known exception 
is water in the temperature range between $0$ and $ + 4$ $^{\circ}$C). 
Parameter $\alpha$ 
---characterizing the ``rapidity'' of increase in temperature with distance from the disk rotation axis---can be either positive or negative. 
When $\alpha > 0$, 
the periphery of the accretion disk is heated more than the central zone; 
when $\alpha < 0$, 
the central zone of the disk is hotter than the periphery. 
Thermal length scale  $\lambda_{\theta} =  ( | \alpha | \beta)^{-1/2}$  
is determined by the background state of the disk in the framework of the model. 
Obviously, space scale parameter $R$ characterizes the ``rapidity'' of the decrease in temperature 
(and stream function) with distance from the disk rotation axis. 
Below, we write $R^2 \rightarrow R^2 \, ( \Theta (\alpha) - \Theta (- \alpha ))$, 
where $\Theta (\alpha)$ is the Heaviside step-function 
(equal to unit for positive argument and zero for negative argument),  
and, leaving the old notations, 
$R^2 = ( | \alpha |  \beta)^{-1} (\omega / \Omega)^2 $.

\section{Results}
\label{s:3} 

We use the  model described  above to gain insights into the three-spot structure in the EHT image of Sgr~A* accretion disk (Figure~\ref{Fig1}A).  
The bright spots are the zones with higher (relative to the base level) temperature $\tau$  
and (since $\tau \sim \psi$) with higher vorticity~$\Delta \psi$. 
In this framework, thermo-hydrodynamical vorticity-field structures may be conceptually analyzed 
with two limit approaches:  
the method of localized vortices  and 
the method of thermo-vorticial spots (blobs).

{\bf Localized Vortices}:
The simplest consideration for a three-spot stationary thermo-vortex structure 
(symmetric with respect to the rotation axis of a pancake-like thin accretion disk) is 
to analytically treat the hot zones as narrowly localized formations modeled as 2D~delta-functions at the limit case. 
Doing so would allow us to find the required conditions for the existence of the observed phenomenon and the resulting relationships between key characteristics of the structure. 
(This approach is not limited to only structures with three spots, but can be generalized to structures with any vortex-number $N$.) 
Hence, we write: 
\begin{eqnarray}
\Delta \psi - R^{-2} ( \Theta (\alpha) - \Theta (- \alpha )) \psi = \sum_{j=1}^{3} \kappa R^2 \, \delta^{(2)} ({\mathbf x} - {\mathbf x}^{(j)}), 
 \label{eq:22} 
\end{eqnarray} 
where $\kappa$ characterizes the intensity of one localized vortex, and
$\delta^{(2)}  ({\mathbf x} - {\mathbf x}^{(j)}) \equiv \delta^{(1)}  (x - x^{(j)}) \times \delta^{(1)}  (y - y^{(j)})$ is the two-dimensional Dirac function in the $xy$-disk plane. 
The appearance of the factor $R^2$ in the right side of Equation~(\ref{eq:22}) is due to the following reasoning: the delta-function is a dimensional function. Its dimension is $[\delta^{(2)}] = [length]^{-2}$ for 2D space. Obviously, the argument of the delta-function must be dimensionless.
(Indeed, there is no such thing as $sin$ of ``one inch'' or $tan$ of ``ten gallons''.) 
Therefore, 
$\delta^{(2)} \equiv \delta^{(1)}  (R^{-1} x - R^{-1} x^{(j)}) \times \delta^{(1)}  (R^{-1} y - R^{-1} y^{(j)}) = 
R^2 \, \delta^{(2)} ({\mathbf x} - {\mathbf x}^{(j)})$. 
In view of the meaning of the delta function in the right part of Equation~(\ref{eq:22}) 
and of the structure of the left part of Equation~(\ref{eq:22}),  
the scale-factor $R$ is the same $R$ that was introduced above---the  
 characteristic size of the vortex kernel. 
 Although alternatives may exist, following the Occam's Razor principle---the 
 problem-solving principle of parsimony that ``entities should not be multiplied beyond 
 necessity''---we choose the simplest option. 
 Indeed, among the options to use $R$ or $R$ multiplied by dimensionless $(\kappa / \omega)^n$, 
 we choose the simple $R$. 
 
We further note that $x^{(j)}$ and $y^{(j)}$ may be expressed 
as components of the complex quantity $l \times \exp ( i  2 \pi (j -1) / 3 )$ in the plane $x + i y$  
(where $l$  is the radius of a circumscribed circle).  

When a vortex is modeled by a delta-function, a weak (logarithmic) singularity appears in 
the distribution of stream function $\psi$ (which is the solution of Equation~(\ref{eq:22})) when $\alpha > 0$.  
If $\alpha < 0$, or a  smoother distribution is adopted for $\alpha > 0$, 
any singularities in $\psi$ function then disappear. 

The  solution to Equation~(\ref{eq:22})  for unbounded space is  
\begin{eqnarray}
\psi = - \frac{\kappa R^2}{2 \pi} \sum_{j=1}^3 Z_0 ( \frac{|{\mathbf x} - {\mathbf x}^{(j)}|}{R} ) \, .
 \label{eq:23}
\end{eqnarray} 
Here, 
$Z_0 (\xi) = J_0 (\xi) \Theta (- \alpha) + K_0 (\xi) \Theta (\alpha)$, 
$J_0 (\xi)$ is the Bessel function of order $n=0$, 
$K_0 (\xi)$ is the modified Bessel (McDonald) function of order $n=0$. 
 
 Recall that for small values of the argument,  
 function $J_0 (\xi) \rightarrow 1$,  
 function $K_0 (\xi)$ behaves logarithmically: 
 $K_0 (\xi) \simeq - \log \xi + 0.1156$. 
For large values,  function $J_0 (\xi)$, as known, 
behaves as $J_0 (\xi) \simeq \sqrt {2 / \pi \xi} \cos [ - \xi + \pi / 4 ] $, 
function $K_0 (\xi)$ behaves as $K_0 (\xi) \simeq \sqrt {\pi / (2 \,\xi)} \exp{(-\xi)}$. 
This means that for $\alpha > 0$ and $r \gg R$; i.e., when $\xi \sim r/R \gg 1$, 
function $K_0 (\xi)$ exponentially quickly tends to zero, 
and therefore, the stream function $\psi$, and consequently the local vortex magnitude and temperature excess  $\tau$, all will also tend to zero. 
This is the consequence of the initial choice to model the vortices via delta-functions. 
Numerically, for some characteristic points, $K_0 (1) \simeq 0.421$ and $K_0 (0.4569) \simeq 1$.  
The derivative of $Z_0$ with respect to argument is $Z_0 ' (\xi) = - Z_1 (\xi)$.

By substituting  
Equation~(\ref{eq:22})  into Equation~(\ref{eq:21}), 
and then taking into account \linebreak \mbox{Equation~(\ref{eq:23}), }  
and  by setting the terms with delta function and their derivatives as equal to zero, 
we find the set of conditions for the existence of the modeled vortex structure: 
\begin{eqnarray}
 \epsilon_{i k}  {\sum_{n=1}^{3} } \bigg( -\omega x_i^{(n)}  + {\sum_{m=1}^3} {'}  \frac{\kappa R^2}{2 \pi R} \frac{( x_i^{(n)} - x_i^{(m)} ) }{| z^{(nm)} |} Z_1 (\frac{| z^{(nm)} |}{R} \bigg)  \kappa R^2 \partial_k \delta  ( z^{(n)} ) = 0 \, .  
 \label{eq:24}
\end{eqnarray} 
In this expression, symbol ``prime'' in the second summation indicates that the effect of vortex self-action is excluded. 
When notation $z^{(j)} = {\mathbf x} - {\mathbf x}^{(j)}$ is used,  
then $z^{(ij)} = {\mathbf x}^{(i)} - {\mathbf x}^{(j)}$ and 
$| z^{(12)} |= | z^{(23)} | = | z^{(31)} | = \sqrt{3}\,  l$  (where $l$  is the radius of a circumscribed circle). 

To satisfy Equation~(\ref{eq:24}), the coefficients before every singularity 
must equal  zero. 
Thus, the following condition for the parameters of the vortex configuration arises: 
\begin{eqnarray}
  -\omega l^2  +  \frac{2 \kappa R^2}{2 \pi R} \frac{ l^2 (1 - \cos 2 \pi / 3 )}{l \sqrt{3}} Z_1 (\frac{\sqrt{3} l }{R} ) = 0 \, .  
 \label{eq:25}
\end{eqnarray} 
\textls[-15]{Recall that
$R = (| \alpha | \beta)^{-1/2} | \omega / \Omega) |$ 
is a dimensional function of four dimensional parameters. }

\begin{figure}[H] 
\begin{minipage}{0.42\textwidth}
     \includegraphics[width=0.99\textwidth]{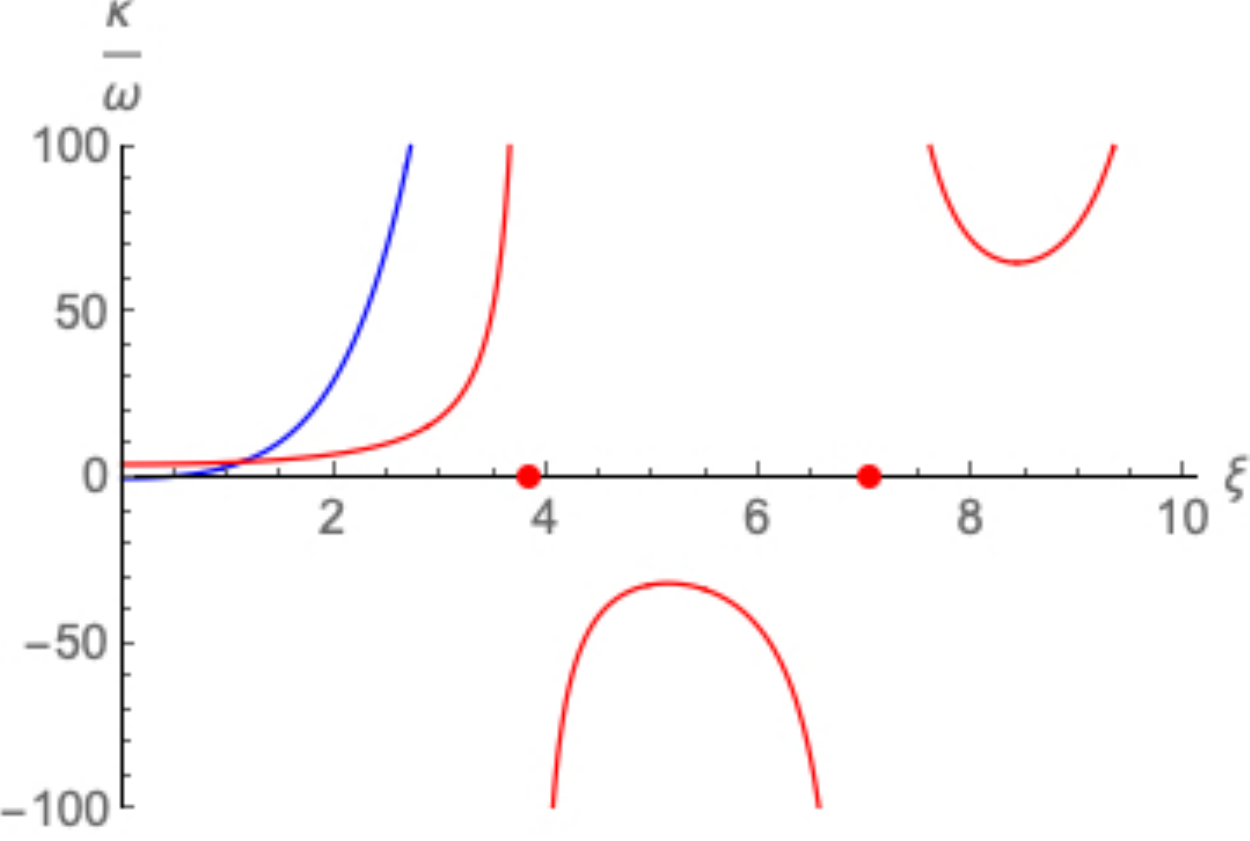}
 \end{minipage}
 \begin{minipage}{0.42\textwidth}
 \centering
        \includegraphics[width=0.99\textwidth]{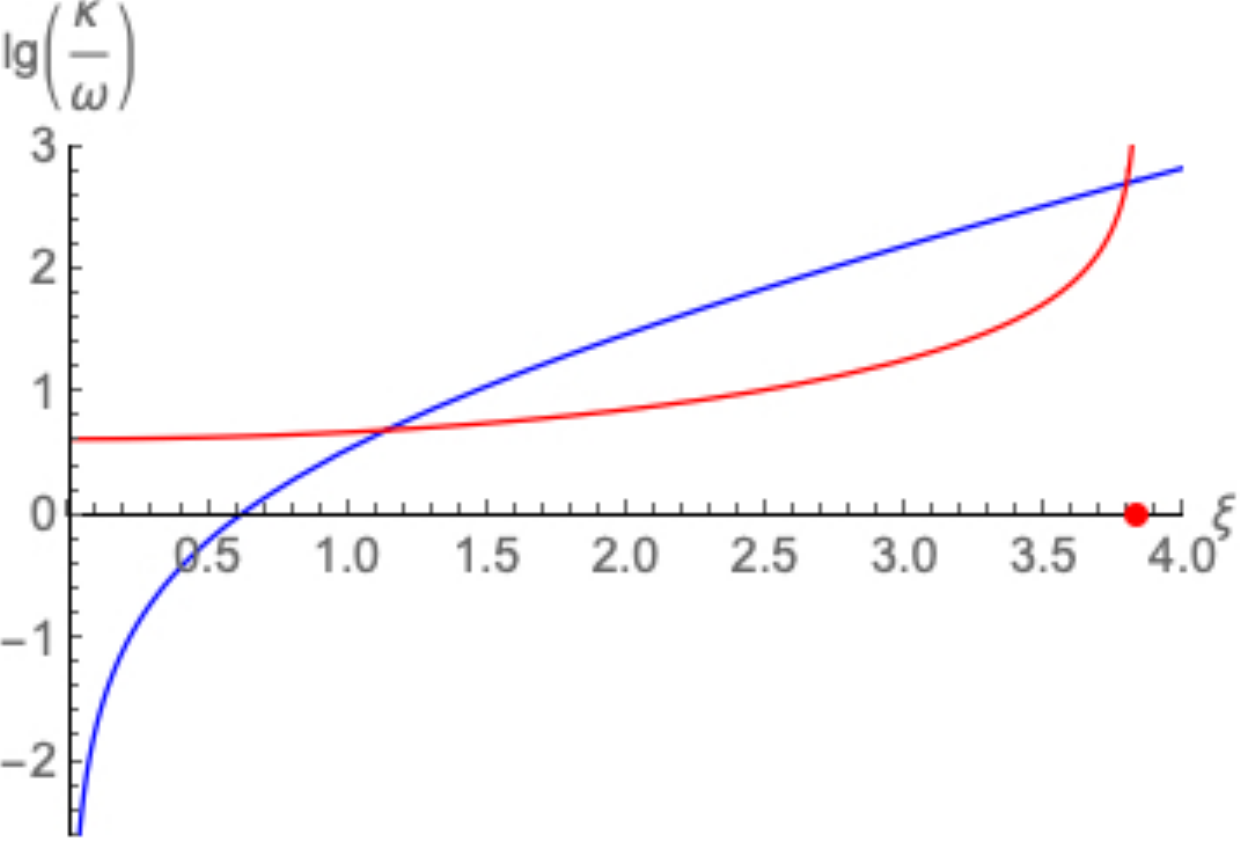}
 \end{minipage} 
\caption{
Model of Localized Vortices.  
{\bf Left panel~(A):}  Both axes are in linear scale. 
{\bf Right panel~(B):}  Vertical axis is in log-scale.
{\bf Both panels~(A,B):}   
A structure with three vortices forms only when $\kappa/\omega = ( 2 \pi / 3)  ( \xi / Z_1 (\xi) )$, 
where $\xi = \sqrt{3} (l/R)$,  
$l$ is the radius of a circumscribed circle, 
$R = (| \alpha | \beta)^{-1/2} | \omega / \Omega) |$,  
$\alpha$ is the parameter characterizing the ``rapidity'' of increase in temperature with distance from the disk rotation axis, 
$\beta = - \rho^{-1} (\partial \rho / \partial \theta )$ is  the coefficient of thermal expansion, 
$\omega$ is the angular velocity of rotation of the vortex structure,  
$\Omega$ is the angular velocity of rotation of the accretion disk, and 
$\kappa$ is the characteristic intensity of one localized vortex.   
The blue curve is for $\alpha > 0$ (the disk periphery is hotter).
The red curve is for $\alpha < 0$  (the disk periphery is cooler):  
 intensity $\kappa \rightarrow \infty$ 
 as $\xi \equiv \sqrt{3} (l/R) \rightarrow \xi_N$ (red dot) at which the Bessel function $J_1 (\xi_N) = 0$. 
(Here $\xi_1 \simeq 3.8317$.)
}
\label{Fig3}
\end{figure}

Equation~(\ref{eq:25}) can be rewritten in a compact dimensionless form, 
where $\xi = \sqrt{3} (l/R)$:   
\begin{eqnarray}
 \frac{\kappa}{\omega} = \frac{2 \pi}{3}  \frac{\xi}{ Z_1 (\xi) } \,     
 \label{eq:26}
\end{eqnarray} 
which says that for any specific $\xi$,  ratio $\kappa / \omega$ is uniquely defined. 
Figure~\ref{Fig3} plots 
\mbox{Equation (\ref{eq:26}).}  

For the presented model, the temperature excess $\tau$ follows from Equation~(\ref{eq:20}) and Equation~(\ref{eq:23}):    
\begin{eqnarray}
 \tau =  \frac{1}{2 \pi \beta } (\frac{\kappa}{\omega}) (\frac{\omega}{\Omega})^2\sum_{j=1}^3 Z_0 ( \frac{|{\mathbf x} - {\mathbf x}^{(j)}|}{R} )   .
 \label{eq:27}
\end{eqnarray} 
The characteristic temperature scale is defined thus as $\tau_0 = (2 \pi \beta)^{-1} (\kappa / \omega) (\omega / \Omega)^2$. 
When $\alpha < 0$ (disk periphery is cooler), 
both ``anticyclones'' (temperatures excesses) and ``cyclones'' (temperature depressions) are possible.

Figure~\ref{Fig1}B (on the front page of the article) illustrates the temperature distribution for a system with three localized vortices, calculated using Equation~(\ref{eq:27}), for a case when $\alpha > 0$ (disk periphery is hotter). 
In Figure~\ref{Fig1}B, 
the unit for the color scale  
is the characteristic magnitude of temperature excess 
$\tau_0 = (2 \pi \beta)^{-1} (\kappa / \omega) (\omega / \Omega)^2$. 
For the $x$ and $y$ axes, the unit is 
$R = \lambda_{\theta} (\omega / \Omega)$;    
$\omega$ is the angular velocity of rotation of the vortex structure as a whole;    
$\Omega$ is the characteristic angular velocity of rotation of the accretion disk; 
and 
$\lambda_{\theta}= (\alpha \beta)^{-1/2}$. 
In the expression for $\lambda_{\theta}$, 
the parameter  $\alpha$ 
(characterizing the ``rapidity'' of the increase in temperature with distance from the disk rotation axis) 
may be estimated as 
$\alpha \sim (\Delta \theta / R_*^2)$. 
Here,  
$R_*$ is 
the characteristic external radius  of the active part of the accretion disk 
(obviously not equal to $R$) is presumed to be greater than the size of the macrostructure,  
and 
 $\Delta \theta$ may be estimated as the 
 temperature difference across the disk (from the periphery to the center). 
In the expression for $\lambda_{\theta}$, 
the parameter $\beta = - \rho^{-1} (\partial \rho / \partial \theta )$ is  the coefficient of thermal expansion, 
which (for the model whose equation of state is approximated by the equation for an ideal gas)  
may be written as 
$\beta \simeq \theta^{-1}$, where   
$\theta$ is the averaged temperature of the accretion disk.

\begin{figure}[H]
\centering 
\begin{minipage}{0.27\textwidth} 
\includegraphics[width=0.79\textwidth]{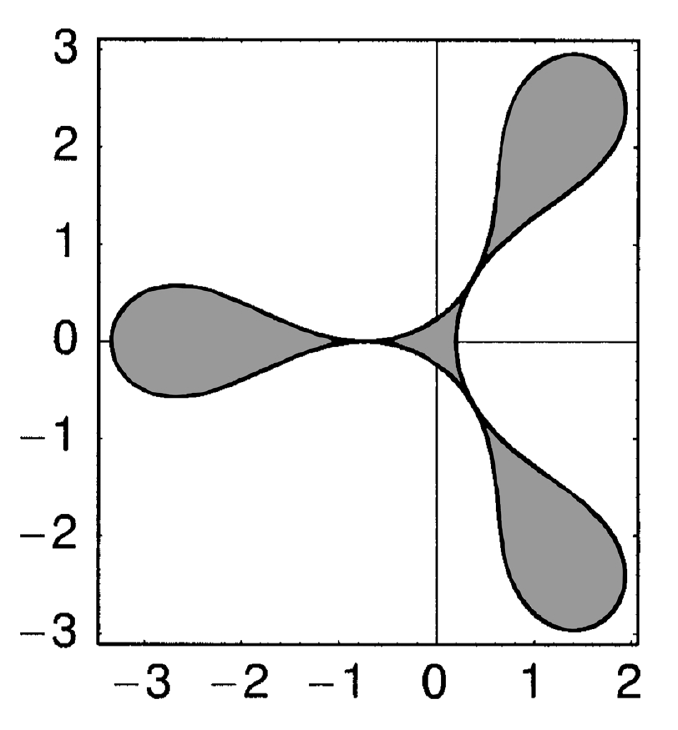} 
\end{minipage}
\begin{minipage}{0.44\textwidth} 
\includegraphics[width=0.79\textwidth]{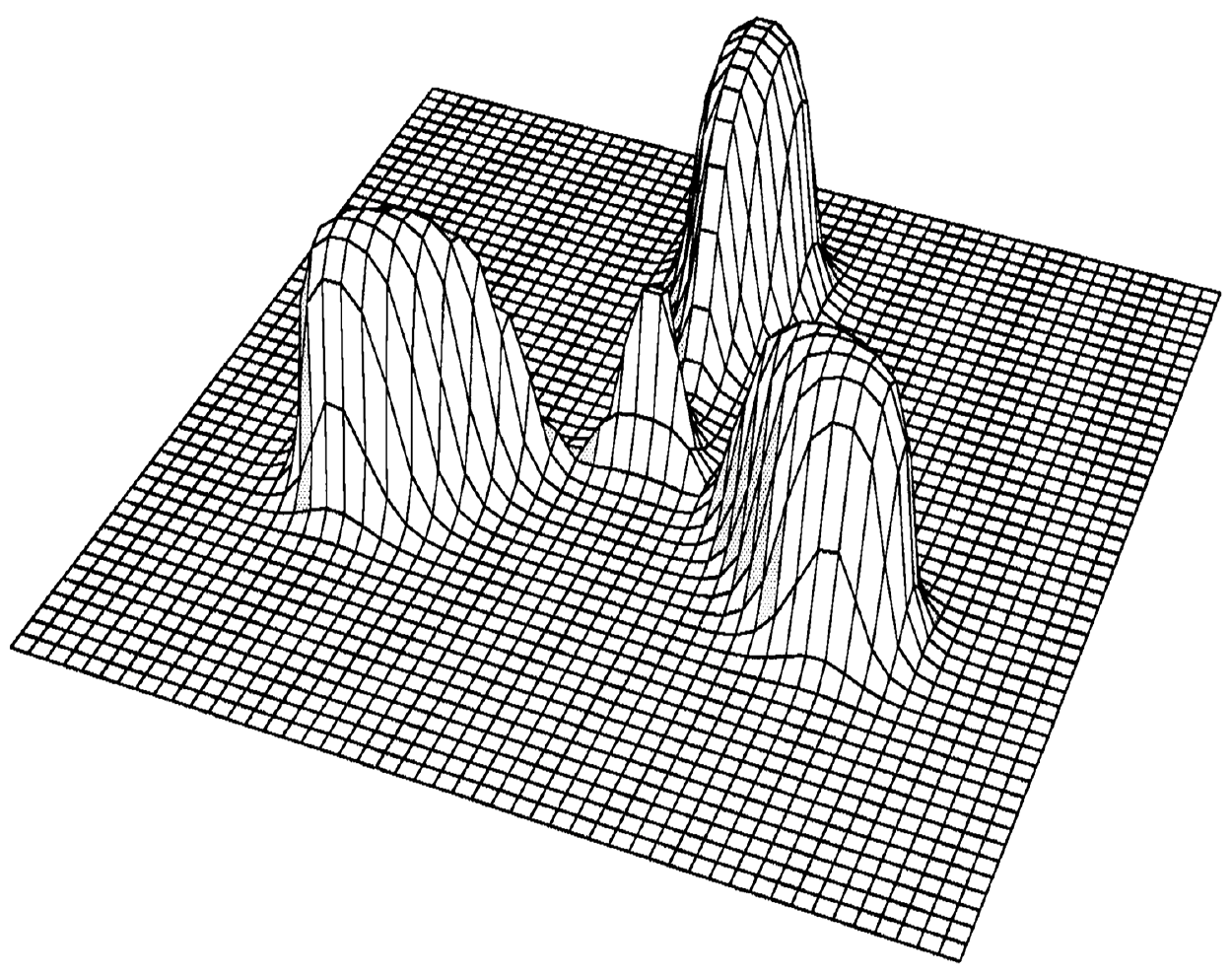} 
\end{minipage}
\caption{
Illustration of analysis via method of thermo-vorticial spots (``blobs''). 
{\bf Left panel (A):}  
A three-petal vorticity distribution is specified as 
$q ({\mathbf x}) = q_0 \Theta (x,y)$, where  
vorticity $q_0$ is constant and 
function $\Theta (x,y)$ is two-dimensional Heaviside step-function 
equal to one inside the shaded domain and zero outside. 
(Ref.~\cite{GonPav2008} 
explains in detail the entire theory and methodology. 
This depicted three-petal thermo-vorticial structure 
takes this particular shape when one of the guiding parameters of the macro-configuration reaches its limit case.)  
{\bf Right Panel (B):}   
For the vorticity distribution specified in Panel~(A),  
the resulting 
temperature distribution $\tau ({\mathbf x})$ is 
obtained per Equation~(\ref{eq:29}).  
The unit along the vertical axis is the characteristic magnitude of temperature excess 
$\tau_0 =  q_0 \alpha R L / 2 \pi \omega$.  
(The central $\tau$ peak is a consequence of the vorticity parametrization for the geophysical application in Ref.~\cite{GonPav2008}, which for a black hole accretion disk should be obviously nil.)
{\bf Both Panels (A,B):} For $x$ and $y$ axes, the unit is 
$R = \lambda_{\theta} (\omega / \Omega)$;    
$\omega$ is the angular velocity of rotation of the vortex structure as a whole;    
$\Omega$ is the characteristic angular velocity of rotation of the accretion disk; 
and 
$\lambda_{\theta}= (\alpha \beta)^{-1/2}$. 
In the expression for $\lambda_{\theta}$,  the
parameter  $\alpha$ 
(characterizing the ``rapidity'' of increase in temperature with distance from the disk rotation axis) 
may be estimated as 
$\alpha = (8/ \pi^2) (\Delta \theta / R_*^2)$. 
Here  
$R_*$ is 
the characteristic radius  of the accretion disk 
(obviously not equal to $R$) is presumed to be greater than  size of the macro-structure,  
and 
 $\Delta \theta$ may be estimated as the 
 temperature difference across the disk (from the periphery to the center). 
In the expression for $\lambda_{\theta}$, the 
parameter $\beta = - \rho^{-1} (\partial \rho / \partial \theta )$ is  the coefficient of thermal expansion, 
which (for the model whose equation of state is approximated by the equation for an ideal gas)  
may be written as 
$\beta \simeq \theta^{-1}$, where   
$\theta$ is the averaged temperature of the accretion disk.
}
 \label{Fig4}
\end{figure}

{\bf Thermo-Vorticial Spots (``Blobs'')}:
Another fruitful approach is to 
use the concept of thermo-vortical spots, 
for which the 
vorticity is constant inside domains bounded by movable boundaries (contours) and 
is zero outside.  
Then,  the problem of vortex evolution becomes the problem of examination of the movement of the contours and determination of their final macro-configuration. 
Examples of such macro-configurations are depicted in Figures~\ref{Fig4} and~\ref{Fig5} (discussed below). 

Within this framework---called the contour dynamics method (CDM) (see \linebreak
Refs.~%
\cite{GonPav1998b,GonPav2008,GonPav2000,GonGrPav2002,GonPav2001,GonPav2001b,GonPav2003})---the 
distribution of vorticity $q$ for a vortex structure may be 
written as $q = q_0 \Theta (x,y)$. 
Function $\Theta (x,y)$ is a two-dimensional Heaviside step-function 
equal to one inside the domain in consideration and zero outside. 
Parameter~$q_0$ describes the vorticity inside the spot, which is presumed constant. 
Obviously, the dimension of this parameter is $[q_0] = [time]^{-1}$.

Using CDM
(see the physical foundation and the 
details on how to perform calculations using CDM in Ref.~\cite{GonPav2008}), 
we find the general expression for temperature distribution: 
\begin{eqnarray}
 \tau (x,y) = - (\frac{q_0 \alpha R L}{2 \pi \omega})  \int_{\partial D} ds \bigg( K_1 (\frac{|\hat{z} - z|}{\mu}) - \frac{\mu}{|\hat{z} - z|} \bigg) \frac{\hat{z}_{,s} (\hat{z} - z^*)}{|\hat{z} - z|} \, .
 \label{eq:29}
\end{eqnarray} 
Here, 
the integral is taken along the contour (whose shape is previously found by solving the problem of the  spot configuration); 
$z = (x + i y)/L$ is the complex variable of the $xy$-position, 
normalized by the characteristic space-scale 
$L = (R / 2) (q_0 / \omega)^{1/3}$, which is linked to the size of the macro-configuration; 
parameter $\mu = R / L$ is dimensionless; and
domain $D$ in the complex plane $z$ 
is a region filled with the vorticity $q_0$. 
The domain is bounded by a closed contour $\partial D$; 
this boundary is expressed in the parametric form $z = \hat{z} (s)$;  
parameter~$s$ is the contour arc length beginning from some initial point; and
 derivative $\hat{z}_{, s} \equiv \partial \hat{z}  / \partial s$ is a unit vector that is 
 obviously tangential to the contour $\partial D$. 
 
Based on the  contour-integral Equation~(\ref{eq:29}), 
the temperature distribution $\tau (x) / \tau_0$  is solved (and plotted in Figure~\ref{Fig4}B) 
for the contour boundary (black line)  in Figure~\ref{Fig4}A. 
Shaded gray in Figure~\ref{Fig4}A is the constant vorticity distribution $q({\mathbf x}) = q_0$.

The key parameters here are 
$\tau_0 =  q_0 \alpha R L / 2 \pi \omega, \; 
R = \lambda_{\theta} (\Omega / \omega), \; 
\lambda_{\theta}= (\alpha \beta)^{-1/2}, \; 
\beta= \theta^{-1} , \; 
 (q_0 / \omega) = 8 (L / R)^3$ 
and $\alpha = (8/ \pi^2) (\Delta \theta / R_*^2)$,  
where 
$\theta$ is the averaged temperature of the accretion disk, 
$R_* (\neq R)$ is the characteristic radius of the accretion disk, and 
 $\Delta \theta$ is the 
 temperature difference across the disk (from the periphery to the center). 
 Thus, the characteristic level of the temperature excess in 
 this limit case (i.e., in the model of thermo-vorticial spots, not in the model of localized thermo-vortices) can be estimated by the expression
 \begin{eqnarray}
 \tau_0 =  \frac{1}{4 \pi \beta} (\frac{q_0}{\omega})^{{4}/{3}} (\frac{\omega}{\Omega})^2 \, .
 \label{eq:30}
 \end{eqnarray} 
The above-mentioned parameters in Equation~(\ref{eq:30}) are 
the very parameters that must be measured experimentally  
to be able to comprehend the phenomena in images 
such as Figure~\ref{Fig1}A. 

Equation~(\ref{eq:30}) is functionally similar to Equation~(\ref{eq:27}). 
The difference is in the index of power dependence of the ratio of the vortex intensity and its angular velocity. 
This is an insignificant difference considering the fact that the two models describe 
radically different limit cases.

\section{Conclusions}
\label{s:4} 

Large-scale long-lived vortices are found in many types of hydrodynamic flows. Large vortices in turbulent flows are called coherent structures. They are observed, for example, in the planetary atmospheres and in the oceans. 
The horizontal scales of the vortices are much greater than the atmospheric or oceanic thicknesses. 
In the simplest geophysical context, examples of such structures are the Gulf Stream rings, the vortices shed from coastal currents, the cyclones and anti-cyclones, the Antarctic Polar Vortex, etc. A very well-known example is Jupiter’s Great Red Spot: a huge vortex plunged in the equatorial flow, which has persisted for more than three centuries;  the presence of  intense small-scale turbulence around it does not destroy it.  
This prompts the question: under what conditions do large-scale vortex structures form? 

In   traditional 3D space, the turbulent motion is usually considered to be homogeneous and isotropic. Common sense and the  laws of thermodynamics show that it is very difficult to extract energy from a fully chaotic system, and only with some additional specific properties of such systems is this possible to realize. Homogeneous, isotropic turbulence, which does not possess any preferred directions or preferred scales, is extremely symmetric, giving birth to large-scale vortices; self-organization seems to be quasi-improbable in this case. It is evident that the breaking of some structural symmetry is one of the necessary conditions for the possibility of self-organization. It becomes clear that turbulent motion of fluid/gas with broken spherical symmetry 
can be a candidate for a system where self-organization of 2D turbulence into large-scale 2D vortex structures can take place.  

Notably, quasi-2D 
flows in ``thin'' pancake-like accretion disks 
(see, e.g., \cite{Liu2022,Armitage2022} on the physics of accretion), 
where  components of hydrodynamical flows that are perpendicular to the disk plane are strongly suppressed, can be a place where large-scale vortices can be self-organized. 
However, another condition is an insignificant influence of a dissipative process on  large-scale motions in the system. As is known, the role of dissipation in a typical hydrodynamical process is characterized by the so-called number of Reynolds, which is determined as the ratio of magnitude of the inertial term  to the dissipative term in the equation of fluid motion. For ``smooth'' flows described by the models of classical hydrodynamics, the  introduction of such a number is not a problem; however, for flows of plasma, the determination of the predominant mechanism of dissipation is not apparent.  
However, the three-spot structure in the EHT image of Sgr~A* accretion disk (seen in Figure~\ref{Fig1}A) 
is a clear example of self-organization in plasma. 

To examine the observed phenomenon in the Sgr~A* disk  from the perspective of theoretical hydrodynamics, 
we first considered  a simplified thermo-hydrodynamic model that permits   analytical consideration. 
In the model, the vorticity clusters are approximated via delta-functions. 
As the result, we established the condition of existence of a regular thermo-vorticial structure 
(Equation~(\ref{eq:26}); see also Figure~\ref{Fig2}). 
We also spelled out 
the relationships that should take place between, on the one hand, 
the parameters that determine  the 
vortex-structure 
dynamics---each vortex size ($l$), its period of proper revolution 
($2 \pi \omega^{-1}$), and 
temperature excess $\tau$ in the vortex---and, on the other hand, 
the accretion disk characteristics ($\alpha$ and $\beta$, i.e.,  $\lambda_{\theta}$)  and 
the angular velocity of the entire accretion disk $\Omega$.

The necessary conditions for the formation of large quasi-stationary symmetric thermo-vorticial structures in the plasma disk are as follows: 
(1) the accretion disk has to be pancake-like thin (i.e., $h \ll R_*$, where $h$ is the thickness and  $R_*$ is the characteristic radius) and rotate with non-zero angular velocity ($\Omega \neq 0$); and 
(2) the disk temperature has to decrease towards the center (i.e., parameter $\alpha > 0$).
A multi-spot thermo-vortex structure forms only 
when key system parameters fall within the ranges captured by the 
dimensionless relationship: 
for a three-spot structure,  
$\kappa / \omega = ( 2 \pi / 3)  ( \xi / K_1 (\xi) )$,  
where  $\xi = \sqrt{3} (l / \lambda_{\theta}) (\Omega / \omega)$. 
The temperature of the vortices $\tau$ (i.e., the excess over the base level) is also linked to another parameter of the system, vortex intensity $\kappa$ of the hot spot: 
$\tau \sim \kappa$. 
Because the bright spots are hot, i.e., $\tau>0$, this result also means that the 
rotation directions of the vortices $\omega$ and the entire accretion disk $\Omega$ must co-align.
 
In the framework of localized vortices, 
the estimate for temperature excess of the bright hot spots
is given by Equation~(\ref{eq:27}),
and in the framework of  thermo-vorticial spots (with sharp contour boundaries), by  Equation~(\ref{eq:30}).

 \begin{figure}[H] 
     \includegraphics[width=0.5\textwidth]{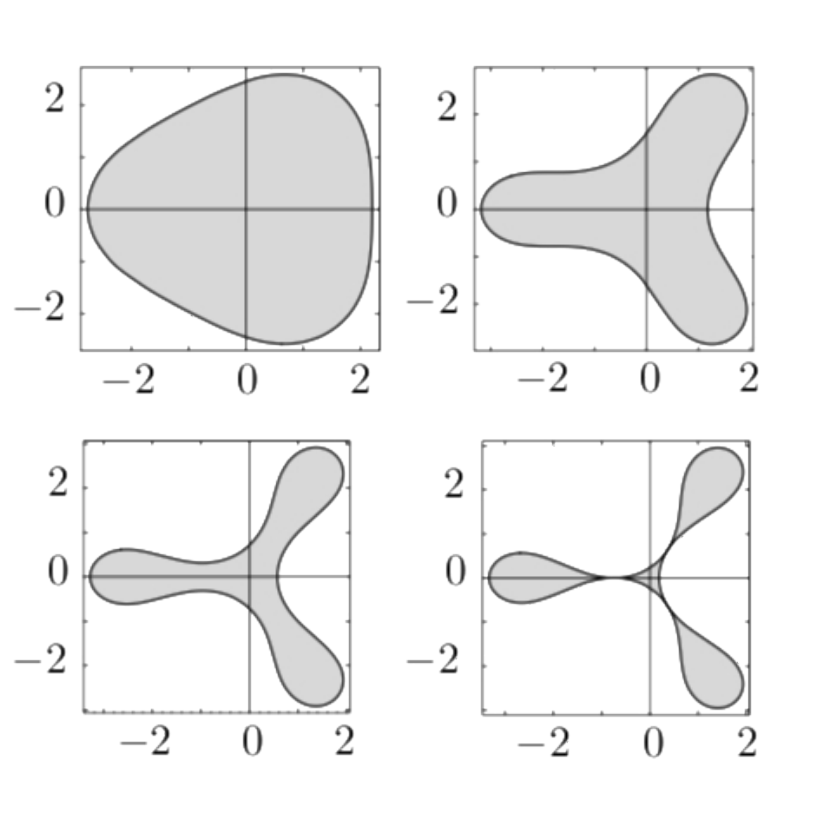}
  \caption{
  The four panels 
  (obtained in Ref.~\cite{GonPav2008} via the method of thermo-vorticial spots) 
  illustrate 
that uniformly rotating 
macro-structures may take 
various shapes 
depending on 
 guiding parameters:  
the shape may range from a weakly deformed circle to a 
sharply pronounced three-petal ``flower''. 
Other guiding parameters define how many petals appear: two, three, etc. 
The theory and models are explained in detail in Ref.~\cite{GonPav2008}. 
  The units for axes are the same as in Figure~\ref{Fig4}. 
  }      
 \label{Fig5}   
 \end{figure}
 
For quasi-2D flows in thin layers of ideal fluid, another 
fruitful approach to analysis also exists (for deeper insight, 
see, Refs.~\cite{GonPav2008,GonPav1998b,GonPav2000,GonGrPav2002,GonPav2001,GonPav2001b,GonPav2003},   
and applications and references therein). 
This approach is called the contour dynamics method (CDM).   
The gist of the method is that 
the  continuous hydrodynamical velocity distribution may be treated as a set of 
patches with movable boundaries (contours) and constant vorticity inside. 
As the underpinnings of the CDM show, such vortex-patch approximation correctly grasps the general tendency in the dynamics and evolution of large-scale flows when the 
large-scale motions are weakly sensitive to a fine structure in the hydrodynamical velocity field. 
The contours of the vortex structures within the CDM framework 
  are determined by  
spatially one-dimensional integro-differential nonlinear equations. 
Equations of the contour dynamics---which describe the self-induced motion of the vorticity-discontinuity boundaries, or “contours”, in an inviscid, incompressible, two-dimensional fluid with piecewise constant vorticity distribution---may be effectively resolved by either numerical or analytical approaches. 
Figure~\ref{Fig5} 
illustrates the variability of shapes that a symmetric stationary macro-structure may take, 
depending on one of its guiding parameters.  
The shape may range from a weakly deformed circle to a 
sharply pronounced three-petal ``flower''. 
(Other parameters define how many petals appear: two, three, etc.) 

 In view of the visual similarity between the hot-spot arrangement in the Sgr~A* accretion disk
 revealed in Figure~\ref{Fig1}A 
 and the thermo-vortex-structure plotted in Figure~\ref{Fig5}, 
 we conclude that the observed hot spots in the Sgr~A* accretion disk are highly likely to be large-scale quasi-2D quasi-stationary {\em vortices} in their nature.
 
In conclusion, let us emphasize that further improvement in understanding 
undoubtedly depends on 
the progress in numerical simulations, for which 
 solid {\em experimental} knowledge of parameters defining the processes is paramount.
 Specifically, as discussed above, these parameters are 
 the background temperature, its contrast,
 the size of the macro-structure,
 the number of petals,
 the angular velocity of rotation of the structure as a whole,
 the angular velocity of rotation of the accretion disk, 
 gradients of velocity of the collective flows of plasma in the disk, and
 the characteristic time of existence of this \mbox{quasi-stationary structure. }

%%%%%%%%%%%%%%%%%%%%%%%%%%%%%%%%%%%%%%%%%%

\vspace{12pt}
\authorcontributions{Conceptualization and Writing, E.P.T., V.P.G. and V.I.P. All authors have read and agreed to the published version of the manuscript.}

\funding{This research received no external funding.}

%\funding{Please add: ``This research received no external funding'' or ``This research was funded by NAME OF FUNDER grant number XXX.'' and  and ``The APC was funded by XXX''. Check carefully that the details given are accurate and use the standard spelling of funding agency names at \url{https://search.crossref.org/funding}, any errors may affect your future funding.}
%\funding{}

%\acknowledgments{In this section, you can acknowledge any support given which is not covered by the author contribution or funding sections. This may include administrative and technical support, or donations in kind (e.g., materials used for experiments).}
%\acknowledgments{}

\dataavailability{Data sharing not applicable.}  %MDPI: We encourage all authors of articles published in MDPI journals to share their research data. In this section, please provide details regarding where data supporting reported results can be found, including links to publicly archived datasets analyzed or generated during the study. Where no new data were created, or where data is unavailable due to privacy or ethical re-strictions, a statement is still required. Suggested Data Availability Statements are available in section “MDPI Research Data Policies” at \url{https://www.mdpi.com/ethics}.

\conflictsofinterest{The authors declare that there is no conflicts of interests regarding the publication of this article.}

%%%%%%%%%%%%%%%%%%%%%%%%%%%%%%%%%%%%%%%%%%
\begin{adjustwidth}{-\extralength}{0cm}

\reftitle{References}

\PublishersNote{}
\end{adjustwidth}

\begin{thebibliography}{999}

\bibitem[{Event Horizon Telescope Collaboration}(2022)]{coll2022a}
Akiyama, K.  
 et al.  [Event Horizon Telescope Collaboration].
\newblock First Sagittarius A* Event Horizon Telescope Results. I. The Shadow of the Supermassive Black Hole in the Center of the Milky Way.
\newblock {\em Astrophys. J. Lett.} {\bf 2022}, {\em 930},~L12, https://doi.org/10.3847/2041-8213/ac6674.

\bibitem[{Event Horizon Telescope Collaboration}(2022)]{coll2022b}
Akiyama, K. 
et al.  [Event Horizon Telescope Collaboration]. 
\newblock First Sagittarius A* Event Horizon Telescope Results. II. EHT and Multiwavelength Observations, Data Processing, and Calibration.
\newblock {\em  Astrophys. J. Lett.} {\bf 2022}, {\em 930},~L13, https://doi.org/10.3847/2041-8213/ac6675.

\bibitem[{Event Horizon Telescope Collaboration}(2022)]{coll2022c}
Akiyama, K. 
 et al. [Event Horizon Telescope Collaboration].
\newblock First Sagittarius A* Event Horizon Telescope Results. III. Imaging of the Galactic Center Supermassive Black Hole.
\newblock {\em  Astrophys. J. Lett.} {\bf 2022}, {\em 930},~L14, https://doi.org/10.3847/2041-8213/ac6429.

\bibitem[{Event Horizon Telescope Collaboration}(2022)]{coll2022d}
Akiyama, K. 
 et al. [Event Horizon Telescope Collaboration].
\newblock First Sagittarius A* Event Horizon Telescope Results. IV. Variability, Morphology, and Black Hole Mass.
\newblock {\em  Astrophys. J. Lett.} {\bf 2022}, {\em 930},~L15, https://doi.org/10.3847/2041-8213/ac6736.

\bibitem[{Event Horizon Telescope Collaboration}(2022)]{coll2022e}
Akiyama, K. 
 et al. [Event Horizon Telescope Collaboration].
\newblock First Sagittarius A* Event Horizon Telescope Results. V. Testing Astrophysical Models of the Galactic Center Black Hole.
\newblock {\em  Astrophys. J. Lett.} {\bf 2022}, {\em 930},~L16, https://doi.org/10.3847/2041-8213/ac6672.

\bibitem[{Event Horizon Telescope Collaboration}(2022)]{coll2022f}
Akiyama, K. 
 et al.  [Event Horizon Telescope Collaboration].
\newblock First Sagittarius A* Event Horizon Telescope Results. VI. Testing the Black Hole Metric.
\newblock {\em  Astrophys. J. Lett.} {\bf 2022}, {\em 930},~L17, https://doi.org/10.3847/2041-8213/ac6756.

\bibitem[{Tito} and {Pavlov}(2018)]{Tito2018}
{Tito}, E.P.; {Pavlov}, V.I.
\newblock Relativistic Motion of Stars near Rotating Black Holes.
\newblock {\em Galaxies} {\bf 2018}, {\em 6},~61, \linebreak https://doi.org/10.3390/galaxies6020061.
  
\bibitem[{Tito} and {Pavlov}(2021)]{Tito2021}
{Tito}, E.P.; {Pavlov}, V.I.
Black Hole Spin and Stellar Flyby Periastron Shift. 
{\em Universe} {\bf 2021}, {\em 7},~364, \linebreak {https://doi.org/10.3390/universe7100364}.
 
\bibitem[{Landau} and {Lifshitz}(2013)]{Landau_Fields}
{Landau}, L.D.; {Lifshitz}, E.M.
{\em The Classical Theory of Fields}; Elsevier: Amsterdam, The Netherlands, 2013.
 
\bibitem[{Misner} {et~al.}(1973){Misner}, {Thorne}, and {Wheeler}]{Misner1973}
{Misner}, C.W.; {Thorne}, K.S.; {Wheeler}, J.A.
{\em {Gravitation}}; W. H. Freeman and Company, Princeton University Press: Princeton, NJ, USA, 1973.
 
\bibitem[{Shapiro} and {Teukolsky}(1983)]{Shapiro1983}
{Shapiro}, S.L.; {Teukolsky}, S.A.
{\em Black Holes, White Dwarfs, and Neutron Stars}; Wiley: Hoboken, NJ, USA, 1983. \linebreak {https://doi.org/10.1002/9783527617661}. 

\bibitem[Visser(2008)]{Visser2007}
Visser, M.
The Kerr Spacetime: A Brief Introduction. 
 {\em arXiv} {\bf 2008}, arXiv:0706.0622v3.
 %Available online: \url{{https://arxiv.org/abs/0706.0622}} (accessed on 20 April 2018).

\bibitem[Frolov and Zelnikov(2011)]{Frolov2011}
Frolov, V.P.; Zelnikov, A.
{\em Introduction to Black Hole Physics}; Oxford University Press: Oxford, UK, 2011.
 
\bibitem[Weinberg(1969)]{Weinberg1972}
Weinberg, S.
\newblock{\em  Gravitation and Cosmology: Principles and Applications of the General Theory of Relativity}; Wiley: New York, NY, USA, 1972.

\bibitem{Kadomtsev1988}
Kadomtsev, B.B. 
\newblock {\em Collective Phenomena in Plasma}, 2nd ed.; USSR:  Moscow, Russia, 1988.
  

\bibitem{LandauLifshitz1987}  
Landau, L.D.; Lifshitz, E.M. 
{\em Fluid Mechanics}, 2nd ed.; rev., 
Pergamon Press: Oxford, NY, USA, 1987.

\bibitem{Turner1979}  
Turner, J.S. 
{\em Buoyancy Effects in Fluids}; Cambridge University Press: Cambridge, UK, 1979.

\bibitem[{Goncharov} and {Pavlov}(1998)]{GonPav1998b}  
{Goncharov}, V.P.; {Pavlov}, V.I. 
Two-dimensional vortex motions of fluid at large Reynolds numbers. 
{\em Phys. Fluids} {\bf 1998}, {\em 10},~2384--2395, {https://doi.org/10.1063/1.869755}

\bibitem[Runge(2021)]{Runge2021}
{Runge}, J.; {Walker}, S.A.  
Probing within the Bondi radius of the ultramassive black hole in NGC 1600. 
 {\em arXiv} {\bf 2021}, arXiv:2102.06216
 %Available online: \url{{https://arxiv.org/abs/2102.06216}} (accessed on 17 December 2022).

\bibitem[{Wong(2011)}(2011)]{Wong2011}
{Wong}, K.-W.; {Irwin}, J.A.; {Yukita}, M.; {Million}, E.T.; {Mathews}, W.G.; {Bregman}, J.N.
\newblock Resolving the Bondi accretion flow toward the supermassive black hole of NGC 3115 with {\em Chandra}. 
\newblock {\em  Astrophys. J. Lett.} {\bf 2011}, {\em 736},~L23, https://doi.org/10.1088/2041-8205/736/1/L23


\bibitem[{Goncharov} and {Pavlov}(1993)]{GonPav1993}
{Goncharov}, V.P.; {Pavlov}, V.I. 
{\em Problemy Gidrodinamiki v Gamil'tonovom Opisanii}; Izd-vo Moskovskogo Universiteta: Moskva, 
 Russia, 1993.

\bibitem[{Goncharov} and {Pavlov}(2008)]{GonPav2008}
{Goncharov}, V. P.; {Pavlov}, V.I.
{\em Hamiltonian Vortex and Wave Dynamics}; Geos: Moscow, Russia, 2008.

\bibitem[{Goncharov} and {Pavlov}(2000)]{GonPav2000}
{Goncharov}, V.P.; {Pavlov}, V.I. 
Large-scale vortex structures in shear flow. 
{\em Eur. J. Mech. B/Fluids} {\bf 2000}, {\em 19},~831, \linebreak {https://doi.org/10.1016/S0997-7546(00)00132-1}
  
\bibitem[{Goncharov}, {Gryanik}, and {Pavlov}(2002)]{GonGrPav2002}
{Goncharov}, V.P.; {Gryanik}, V.M.; {Pavlov}, V.I. 
Venusian “hot spots”: Physical phenomenon and its quantification. 
{\em Phys. Rev. E} {\bf 2002}, {\em 66}, 	066304. {https://doi.org/10.1103/physreve.66.066304}  
  
\bibitem[{Goncharov} and {Pavlov}(2001)]{GonPav2001}
{Goncharov}, V.P.; {Pavlov}, V.I. 
Multipetal Vortex Structures in Two-Dimensional Models of Geophysical Fluid Dynamics and Plasma. 
{\em Soviet J. Exp.  Theor. Phys.} {\bf 2001}, {\em 92},~594--607, {https://doi.org/10.1134/1.1371341} 



\bibitem[{Goncharov} and {Pavlov}(2001)]{GonPav2001b}
{Goncharov}, V.P.; {Pavlov}, V.I. 
Cyclostrophic vortices in polar regions of rotating planets. 
{\em Nonlin. Process.   Geophys.} {\bf 2001}, {\em 8},~301--311, 
{https://doi.org/10.5194/npg-8-301-2001}.


\bibitem[{Goncharov} and {Pavlov}(2003)]{GonPav2003}
{Goncharov}, V.P.; {Pavlov}, V.I.
Hamitonian contour dynamics. 
In {\em Fundamental and Applied Problems of the Vortex Theory};  
{Borisov}, A.V., {Mamaev}, I.S.,  {Sokolovskiy}, M.A.,  Eds.; 
Institute of Computer Science: Moscow and Izhevsk, Russia, 2003; pp.~179--237. (In Russian)

\bibitem[{Liu} and {Qiao}(2022)]{Liu2022}
{Liu}, B.F.; {Qiao}, E.
\newblock Accretion around black holes: The geometry and spectra. 
\newblock {\em iScience} {\bf 2022}, {\em 25},~103544, 
arXiv:2201.06198.
 %Available online: \url{{https://arxiv.org/abs/2201.06198}} (accessed on 19 December 2022).

\bibitem[{Armitage}(2022)]{Armitage2022}
{Armitage}, P.J.
\newblock Lecture notes on accretion disk physics.  \emph{arXiv} \textbf{2022},  
arXiv:2201.07262.
 %Available online: \url{{https://arxiv.org/abs/2201.07262}} (accessed on 19 December 2022).

 \end{thebibliography}
\end{document}